\documentclass[3p,times,procedia]{elsarticle}
\usepackage{nupha_ecrc}

\volume{00}

\firstpage{1}

\journalname{Nuclear Physics A}

\runauth{}

\jid{nupha}

\jnltitlelogo{Nuclear Physics A}

\usepackage{amssymb}

\usepackage[figuresright]{rotating}

\usepackage{bm}

\begin{document}

\begin{frontmatter}



\dochead{}

\title{Hydrodynamics and critical slowing down}


\author[1]{M.~ Stephanov}
\author[2]{Y.~Yin}

\address[1]{Physics Department, University of Illinois at Chicago,
  W. Taylor St., Chicago IL 60607-7059}
\address[2]{Center for Theoretical Physics, Massachusetts Institute of Technology, Cambridge, MA 02139}

\begin{abstract}
We introduce an effective theory which extends hydrodynamics into a
regime where the critical slowing down would otherwise make
hydrodynamics inapplicable.

\end{abstract}


\end{frontmatter}


\section{Introduction}
\label{sec:intro}

Hydrodynamics~\cite{landau2013fluid} is an extremely versatile theory with a wide range of
applications. Its recent developments have largely concentrated on
applications to relativistic heavy-ion collisions where it can
describe bulk evolution of the QCD matter as well as the evolution of
fluctuations and transport of charge, including anomalous chiral
transport. One would like to apply hydrodynamics to
describe the QCD matter evolution near the QCD critical
point~\cite{Stephanov:1998dy}, which would greatly facilitate the
analysis necessary for the discovery of this point in the beam energy
scan experiments. However, hydrodynamics notoriously fails close to
the critical point due to the critical slowing down. The purpose of
this work is to address this shortcoming and propose a solution
\cite{Stephanov-Yin}.

Hydrodynamics is an effective theory describing the dynamical
space-time evolution of the densities of {\em conserved} quantities --
energy, momentum and charge (or charges). The
possibility of such a description is predicated on the separation of
time scales. 
In linearized regime, in terms of the wave vector $k$, the relaxation
rates of hydrodynamic modes is typically proportional to $k^2$ due to the
conservation. All other modes are typically evolving on
much faster scales set by microscopic dynamics, e.g., by collision
rates or temperature.

However, there are important
situations in which there are some modes in the theory which are not
associated with conserved charges, but are nevertheless {\em parametrically}
slow. That means that their evolution rate $\Gamma$ can be made arbitrarily
small by tuning a parameter, even though, for a fixed value of the
parameter,  $\Gamma$ remains {\em finite} in the
hydrodynamic limit $k\to 0$.

In this situation, hydrodynamics is not applicable in the regime where
hydrodynamic modes become almost as fast as the non-hydrodynamic
mode. The domain of applicability of hydrodynamics shrinks to zero as
$\Gamma\to0$. Our goal is to extend the applicability of hydrodynamics
in this context. To achieve this one needs to add an additional
non-hydrodynamic mode associated with the {\em parametrically} slow
relaxation rate~$\Gamma$. We shall refer to such an extension of
hydrodynamics as ``Hydro+''.

Extending hydrodynamics by additional modes is not a new idea. The
most well-known example is the Israel-Stewart theory. However, the
relaxation rate of the additional modes in the Israel-Stewart
extension of the hydrodynamics is on the order of the microscopic
scale and is {\em not} parametrically small. As a result, there is no
justification to include these microscopic modes and leave out others
\cite{Geroch:1995bx}. We wish to explore the situations in which such
a parametric separation of scales between the additional slow mode and
all the other microscopic fast modes makes Hydro+ a well-defined
effective theory.

The two important examples of such additional parametric slowing down
are: 1) The critical slowing down at the critical point; 2) The
hydrodynamics of an axial or chiral charge whose conservation is
violated explicitly by a small parameter, e.g., small quark mass. Both
examples are of direct relevance to the study of the QCD phase diagram
using heavy-ion collisions. 
In
the case of the critical point dynamics, the parameter controlling the
slow-down is the correlation length $\xi$: the characteristic rate of
the relaxation to equilibrium is proportional to $\xi^{-z}$, with
$z\approx 3$ \cite{PhysRevE.55.403}, and can be arbitrarily small as $\xi$ diverges
near the critical point.


A direct indication that hydrodynamic approach breaks
down near the critical point is the divergence of the bulk viscosity $\zeta$,
as $\xi^{z-\alpha/\nu}$, or, in the approximation sufficient for our
discussion, $\xi^3$. The gradient expansion can be
trusted for $c_s k \ll \zeta k^2/w$, which translates to the domain of
applicability of hydrodynamics $k\ll c_sw/\zeta\sim \xi^{-3}$. Near the critical
point this is much less than the inverse correlation length
$\xi^{-1}$, let alone the microscopic scale such as $1/T$.

Such a situation is familiar in effective field theory.  Integrating
fields which are lighter than the scales we are interested in would
lead to the breakdown of locality. To make effective theory {\em local} one
needs to include all the light fields in the effective
description. This is the essence of the Wilsonian paradigm.

Similarly, we can remove the divergent bulk viscosity coefficient and
extend the applicability of hydrodynamics by augmenting it with
an additional mode (or modes) whose relaxation rate vanishes as~$\xi^{-3}$.

\section{Hydro+}

Now, to summarize, we are motivated to consider the following effective
theory describing the evolution of conserved densities plus a
non-hydrodynamic mode we shall call $\phi$. The most important
ingredient of the theory is the non-equilibrium, or more precisely,
quasi-equilibrium entropy $s(\varepsilon,n,\phi)$.
The microscopic meaning of this quantity is, as usual, the logarithm
of the number of the quantum states of the system with given values of
$\varepsilon$, $n$ as well as $\phi$. The derivatives of the entropy
define thermodynamically conjugate quantities $\beta$, $\alpha$ and
$\pi$:
\begin{equation}
  \label{eq:ds}
  ds = \beta d\varepsilon - \alpha dn - \pi d\phi.
\end{equation}
While, due to conservation of energy and charge, $\beta=1/T$ and
$\alpha=\mu/T$ can take arbitrary values in equilibrium, the
equilibrium value of $\pi$ must be zero, since $\phi$ is not a
conserved quantity and will relax to its equilibrium value $\phi_{\rm
  eq}(\varepsilon,n)$, which maximizes the entropy $s(\varepsilon, n,
\phi)$. This relaxation is, however, parametrically slow and allows
us, in a range of time scales, to consider quasi-equilibrium states
characterized by $\phi\neq\phi_{\rm eq}$, or $\pi\neq 0$.

The equations of motion in hydrodynamics, as usual, are the
conservation equations for the stress-energy tensor and the 4-current:
\begin{equation}
  \label{eq:conservation}
  \partial_\mu T^{\mu\nu} =0;\qquad \partial_\mu J^\mu =0.
\end{equation}
Also as usual, to close the system we need to supplement these equations of
motion with the constitutive equations
\begin{equation}
  \label{eq:TmunuJmu}
  T^{\mu\nu} = \varepsilon u^\mu u^\nu + p g_\perp^{\mu\nu}+ \Delta T^{\mu\nu};
\qquad
  J^\mu = n u^\mu + \Delta J^\mu;
\end{equation}
(where $g_\perp^{\mu\nu} = g^{\mu\nu}
 + u^\mu u^\nu$). The pressure $p$ and the kinetic coefficients which appear in the
gradient corrections $\Delta T^{\mu\nu}$ and $\Delta J^\mu$ are
functions of the variables $\varepsilon$, $n$ and $\phi$. 

Finally, Hydro+ must include an additional equation of motion for the additional mode
$\phi$:
\begin{equation}
  \label{eq:phi}
  (u\cdot\partial) \phi = - F_{\phi} - G_\phi(\partial\cdot u),
\end{equation}
where the thermodynamic restoring force $F_\phi$ and the
coefficient $G_\phi$ are given by corresponding constitutive equations
in terms of $\varepsilon$, $n$ and $\phi$ and their gradients.

The second law of thermodynamics $\partial\cdot s\ge0$ imposes
constraints on the constitutive equations. At the leading order (ideal
hydrodynamics) it relates the pressure to the 
entropy and its derivatives:
\begin{equation}
  \label{eq:p}
  \beta p = s - \beta\varepsilon + \alpha n + \pi G_\phi.
\end{equation}
At the lowest nontrivial order in gradients one 
finds that constitutive equations must have the form
\begin{equation}
  \label{eq:J}
  \Delta J_\mu  = -(\lambda\partial_\mu\alpha 
+ \lambda_{\alpha\pi}\partial_\mu\pi);
\end{equation}
\begin{equation}
  \label{eq:Fphi}
  F_\phi = \gamma_\pi \pi 
- \partial_\perp\cdot(\lambda_{\pi\pi}\partial\pi + \lambda_{\alpha\pi}\partial\alpha);
\end{equation}
\begin{equation}
  \label{eq:DeltaTmunu}
  \Delta T^{\mu\nu} = -\eta\sigma^{\mu\nu}
+ \zeta^* g_\perp^{\mu\nu}(\partial\cdot u)
\end{equation}
where the coefficients $\gamma$, $\lambda$ and $\lambda_{\pi\pi}$ as well as the
determinant $\lambda\lambda_{\pi\pi}-\lambda_{\alpha\pi}^2$ must be
non-negative, and so must the viscosities $\eta$ and $\zeta^*$.

Further constraints on the parameters can be obtained using
microscopic derivation of this effective theory, and depend on the
underlying theory that Hydro+ is describing. For example, if $\phi$ is
the axial charge density, the parameter $G_\phi$ equals $\phi$ and in
Eq.~(\ref{eq:p}) we recover a familiar expression for the pressure as
a Legendre transform of the entropy. Parity may also forbid mixing of
axial and vector charges, thus, e.g., $\lambda_{\alpha\pi}=0$. On the
other hand, in the case of the critical point, we find $G_\phi=0$. We
also find that the coefficients $\lambda_{ij}$ are subject to
additional constraints.

The advantage of Hydro+ is that there are no anomalously large kinetic
coefficients which could lead to the breakdown of the gradient
expansion at the scale below $k\sim\xi^{-1}$. Most importantly, the ``bare''
bulk viscosity coefficient $\zeta^*$ in Eq.~(\ref{eq:DeltaTmunu}) does
not diverge as $\Gamma\to0$.

\section{Linearized Hydro+}

\begin{figure}
  \centering
  \includegraphics[width=.5\textwidth]{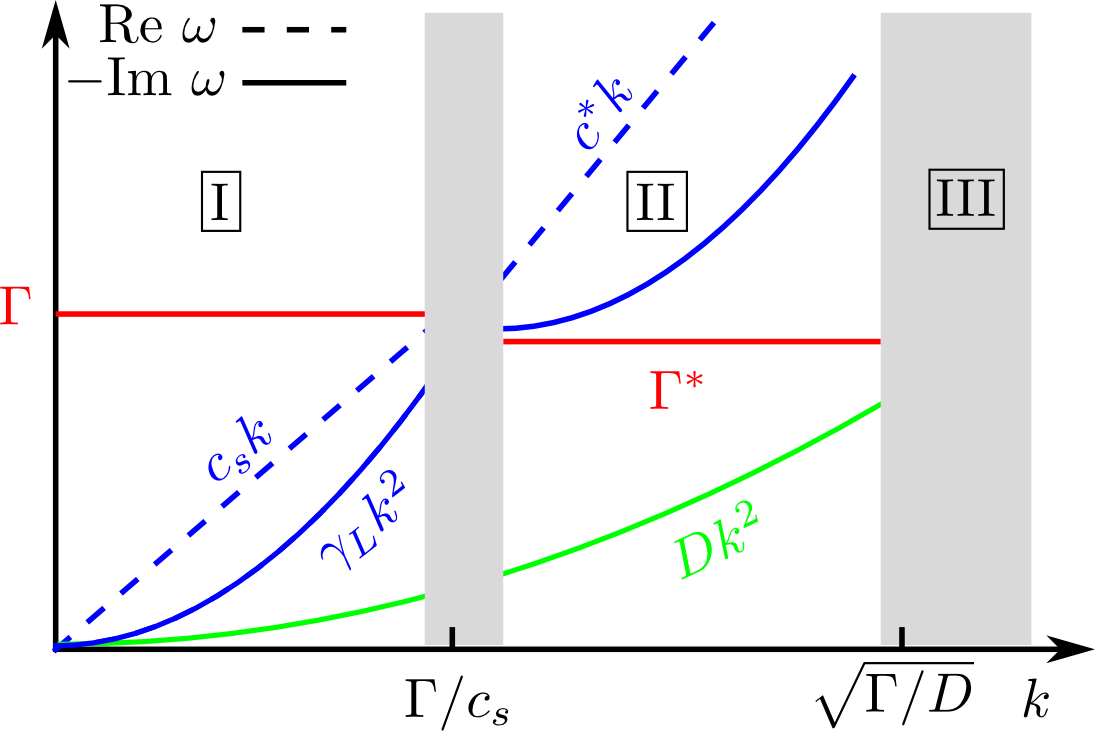}
  \caption{The spectrum of linearized Hydro+. The conventional
    hydrodynamics is valid in Regime I, for as long as the relaxation rate
  $\Gamma$ is faster than the sound oscillation rate. The sound
  attenuation coefficient $\gamma_L\sim\zeta/w\sim c_s^2/\Gamma$ diverges as
  $\Gamma\to0$. In Regime II the sound is the fastest mode and its
  attenuation rate is slower. }
  \label{fig:modes}
\end{figure}

To understand the dynamics and interplay of scales in Hydro+ better it
is useful to consider the spectrum of linearized perturbations around
equilibrium. Figure~\ref{fig:modes} summarizes the results.

As a function of the wave number $k$ one can identify several regimes:

Regime I, or proper hydrodynamic regime. The relaxation rate of the
slow mode $\phi$ is parametrically faster than that of all
hydrodynamic modes (sound and diffusion). The mode effectively
decouples, but as a consequence, the damping coefficient of the sound
mode $\gamma_L\approx \zeta/(2w) \sim c_s^2/\Gamma$ is divergent when
$\Gamma\to0$. The ordinary hydrodynamic regime breaks down at $k\sim
\Gamma/c_s\sim\xi^{-3}$ when the damping rate of the sound is of the
order of the sound frequency and of the order of the relaxation rate
$\Gamma$ (see Fig.~\ref{fig:modes}).

Regime II, or Hydro+ regime. In this regime the sound oscillation rate
is faster than the relaxation rate of the additional slow mode. The
sound speed changes by $\Delta c_s^2 = {c^*_s}^2-c_s^2$ and so does
the rate $\Gamma$, but there is a simple relation between their values
in regimes I and II:
\begin{equation}
  \label{eq:Gammacs}
  \Gamma c_s^2 = \Gamma^* {c^*_s}^2.
\end{equation}
The viscosity coefficients in regimes I and II are related by
\begin{equation}
  \label{eq:zeta}
  \Delta \zeta \equiv \zeta - \zeta^* = \frac{w\Delta c_s^2}{\Gamma}
\end{equation}
which is the relativistic generalization of Landau-Khalatnikov formula
\cite{landau2013fluid}.

Note that the ``bare'' bulk viscosity $\zeta^*$ does not diverge,
while the divergence of the true hydrodynamic bulk viscosity $\zeta$
in Regime I is due to the slow mode rate $\Gamma\to 0$.

Finally, at around $k\sim \sqrt{\Gamma/D}\sim\xi^{-1}$ the relaxation rate of the slow mode, $\Gamma^*$
and the diffusion rate $Dk^2$ reach the same order of magnitude (note
that $D\sim\xi^{-1}$). The dynamics of these slowest modes
should be matched by the mode-coupling dynamics of model-H \cite{RevModPhys.49.435} in
the regime III.

\section{Microscopic origins of Hydro+}

The fluctuations around equilibrium are controlled by the entropy via
Einstein's formula $P\sim e^S$. Near the critical point the entropy as a
function(al) of the hydrodynamic variables has a ``flat direction''. The
direction in which the curvature of the entropy is the smallest near the
critical point is the direction along which the fluctuations are the
largest. It is convenient to rotate the basis of fluctuations from
$\varepsilon$, $ n$ to their linear combinations $\mathcal E$ and
$\mathcal N$, where the flattest direction is $\delta\mathcal E = 0$. The
large fluctuations of the variable $\mathcal N$ can be described by a
two-point function, to which we can apply Wigner transformation:
\begin{equation}
  \label{eq:NN}
  f_{\mathcal N}(t,\bm x,\bm Q) = \int_{\bm y}
\langle\,\delta\mathcal N(t,\bm x + \bm y/2)\,
\delta\mathcal N(t,\bm x - \bm y/2)\,\rangle
e^{-i\bm{Q\cdot y}}.
\end{equation}
We can view $f_{\mathcal N}$ as a mode distribution function, similar
to the particle distribution function in a kinetic theory. The
variable $\bm Q$ can be viewed as an index of the mode. The
characteristic scale for $\bm Q$ is $Q\sim\xi^{-1}$, thus ensuring the
separation of scales between the typical gradients in Hydro+ and the
underlying microscopic theory (model H). The mode distribution
function obeys relaxation equation
\begin{equation}
  \label{eq:f_N-relax}
  (u\cdot\partial)f_{\mathcal N} 
= -\gamma(\bm Q)\, \pi_{\mathcal N}(\bm Q),
\end{equation}
where 
the relaxation coefficient $\gamma(\bm Q)$ is proportional to the
Kawasaki function known from model-coupling theory \cite{PhysRevE.55.403} and
$\pi_{\mathcal N} = \partial s/\partial f_{\mathcal N} $. The
equations of hydrodynamics close once the entropy functional
$s(\varepsilon,n,f_{\mathcal N})$ is given~\cite{Stephanov-Yin}. The resulting theory can
be shown to reproduce the expressions for the frequency-dependent bulk
viscosity and the shift of the sound speed $\Delta c_s^2$.

Hydro+ --- the hydrodynamic theory augmented by a slow mode
$\phi$ or a family of modes $f_{\mathcal N}$ --- should allow
efficient modeling of the evolution of the fireball created in
relativistic heavy-ion collisions, including the regimes where the
critical slowing down and divergent bulk viscosity would otherwise render
conventional hydrodynamic description inapplicable.

This material is based upon work supported by the U.S. Department
of Energy, Office of Science, Office of Nuclear Physics,
within the framework of the Beam Energy Scan Theory (BEST) 
Topical Collaboration and grants Nos. DE-FG0201ER41195 and DE-SC\mbox{0011090}.





\bibliographystyle{elsarticle-num}
\bibliography{hydro+}







\end{document}